\PassOptionsToClass{10pt}{revtex4-1}
\documentclass[aps,prl,showpacs,floatfix,twocolumn,byrevtex,superscriptaddress]{revtex4-1}

\usepackage{amsmath}
\usepackage{amssymb}
\usepackage{amstext}
\usepackage{amsopn}
\usepackage{amsfonts}
\usepackage{amsxtra}
\usepackage[english]{babel}
\usepackage{graphicx}
\usepackage{bm}
\usepackage{multirow}
\usepackage{dcolumn}
\usepackage{color}
\usepackage{hyperref}
\usepackage{todonotes}
\usepackage{verbatim}
\usepackage{soul}

\def\LFA{LiFeAs}
\def\FTS{FeTe$_{0.55}$Se$_{0.45}$}

\begin{document}

\title{\boldmath Decoupled Pairing Amplitude and Electronic Coherence in Iron-Based Superconductors} 
\author{H. Miao}\email[]{hmiao@bnl.gov}
\affiliation{Beijing National Laboratory for Condensed Matter Physics, Institute of Physics, Chinese Academy of Sciences, Beijing 100190, China}
\affiliation{Condensed Matter Physics and Materials Science Department, Brookhaven National Laboratory, Upton, New York 11973, USA}
\author{W. H. Brito}
\affiliation{Condensed Matter Physics and Materials Science Department, Brookhaven National Laboratory, Upton, New York 11973, USA}
\author{Z. P. Yin}
\affiliation{Department of Physics and the Center of Advanced Quantum Studies, Beijing Normal University, Beijing 100875, China}
\author{R. D. Zhong}
\affiliation{Condensed Matter Physics and Materials Science Department, Brookhaven National Laboratory, Upton, New York 11973, USA}
\author{G. D. Gu}
\affiliation{Condensed Matter Physics and Materials Science Department, Brookhaven National Laboratory, Upton, New York 11973, USA}
\author{P. D. Johnson}
\affiliation{Condensed Matter Physics and Materials Science Department, Brookhaven National Laboratory, Upton, New York 11973, USA}
\author{M. P. M. Dean}
\affiliation{Condensed Matter Physics and Materials Science Department, Brookhaven National Laboratory, Upton, New York 11973, USA}
\author{S. Choi}
\affiliation{Condensed Matter Physics and Materials Science Department, Brookhaven National Laboratory, Upton, New York 11973, USA}
\author{G. Kotliar}
\affiliation{Condensed Matter Physics and Materials Science Department, Brookhaven National Laboratory, Upton, New York 11973, USA}
\affiliation{Department of Physics and Astronomy, Rutgers University, Piscataway, New Jersey 08854, USA}
\author{W. Ku}
\affiliation{Physics Department, Shanghai Jiaotong University, 800 Dongchuan Road, Shanghai, 200240, China}
\author{X. C.  Wang}
\affiliation{Beijing National Laboratory for Condensed Matter Physics, Institute of Physics, Chinese Academy of Sciences, Beijing 100190, China}
\author{C. Q.  Jin}
\affiliation{Beijing National Laboratory for Condensed Matter Physics, Institute of Physics, Chinese Academy of Sciences, Beijing 100190, China}
\author{S. -F. Wu}
\affiliation{Beijing National Laboratory for Condensed Matter Physics, Institute of Physics, Chinese Academy of Sciences, Beijing 100190, China}
\author{T. Qian}
\affiliation{Beijing National Laboratory for Condensed Matter Physics, Institute of Physics, Chinese Academy of Sciences, Beijing 100190, China}
\author{H. Ding}
\email[]{dingh@iphy.ac.cn}
\affiliation{Beijing National Laboratory for Condensed Matter Physics, Institute of Physics, Chinese Academy of Sciences, Beijing 100190, China}
\affiliation{Collaborative Innovation Center of Quantum Matter, Beijing 100190, China}

\date{\today}
\begin{abstract}
Here we use angle-resolved photoemission spectroscopy to study superconductivity that emerges in two extreme cases, from a Fermi liquid phase (\LFA{}) and an incoherent bad-metal phase (\FTS{}). We find that although the electronic coherence can strongly reshape the single particle spectral function in the superconducting state, it is decoupled from the maximum superconducting pairing amplitude, which shows a universal scaling that is valid for all FeSCs. Our observation excludes pairing scenarios in the BCS and the BEC limit for FeSCs and calls for a universal strong coupling pairing mechanism for the FeSCs.    
\end{abstract}


\maketitle

The interplay between superconductivity and its normal state electronic coherence remains a central puzzle in unconventional superconductors. In the cuprate and heavy Fermion superconductors, superconductivity emerges from a non-Fermi liquid normal state with nearly vanishing coherent weight, $Z_{k}\rightarrow{}0$, and thus motivated theoretical proposals of superconducting (SC) pairing mechanisms beyond the BCS paradigm \cite{Norman2003, Anderson2004, Lee2006,Garg2008, Keimer2015}. In the multi-orbital iron-based superconductors (FeSCs), the electronic structure and the total carrier density are highly sensitive to the Hund's coupling and the height of anion atoms (As/Se) that are alternatively placed above and below the iron-plane \cite{Yin2012, Kurita2011}. As a consequence, FeSCs display diverse phase diagrams that ignite extensive debates on the pairing mechanism mainly among BCS-like theories that utilize coherent quasi-particles (QPs) near the Fermi level \cite{Mazin2008, Kuroki2008, Kontani2010, Scalapino2012, Hirschfeld2016}, scenarios that emphasize localized electrons with large short-ranged antiferromagnetic (AFM) interactions \cite{Seo2008, Si2008, Yildirim2008, Hu2012, Davis2013}, and strong coupling approach based on metallic continuum and spin fluctuations \cite{Yin2014}. In this paper, we use angle-resolved photoemission spectroscopy (ARPES) to directly explore the evolution of the single-particle spectral function, $A(k,\omega)$, starting from two different phases: (i) a coherent Fermi-liquid phase with large carrier density in \LFA{} and (ii) an incoherent bad metal phase with small carrier density in \FTS{}. We find that while the change of $A(k,\omega)$ in the SC phase strongly depends on $Z_{k}$, superconductivity itself is very robust and shows a universal scaling $2\Delta_{SC}^{max}(k)/k_{B}T_{c}\sim7.2$ for all FeSCs, where $\Delta_{SC}^{max}(k)$ is the maximum SC gap in momentum space determined by ARPES. The independence of $2\Delta_{SC}^{max}(k)/k_{B}T_{c}$ on the correlations and $Z_{k}$ that vary significantly through different families, excludes pairing scenarios in the BCS and the BEC limit and calls for a unified theory for the iron-pnictides and chalcogenides.

%
\begin{figure*}[tb]
\includegraphics[width=16 cm]{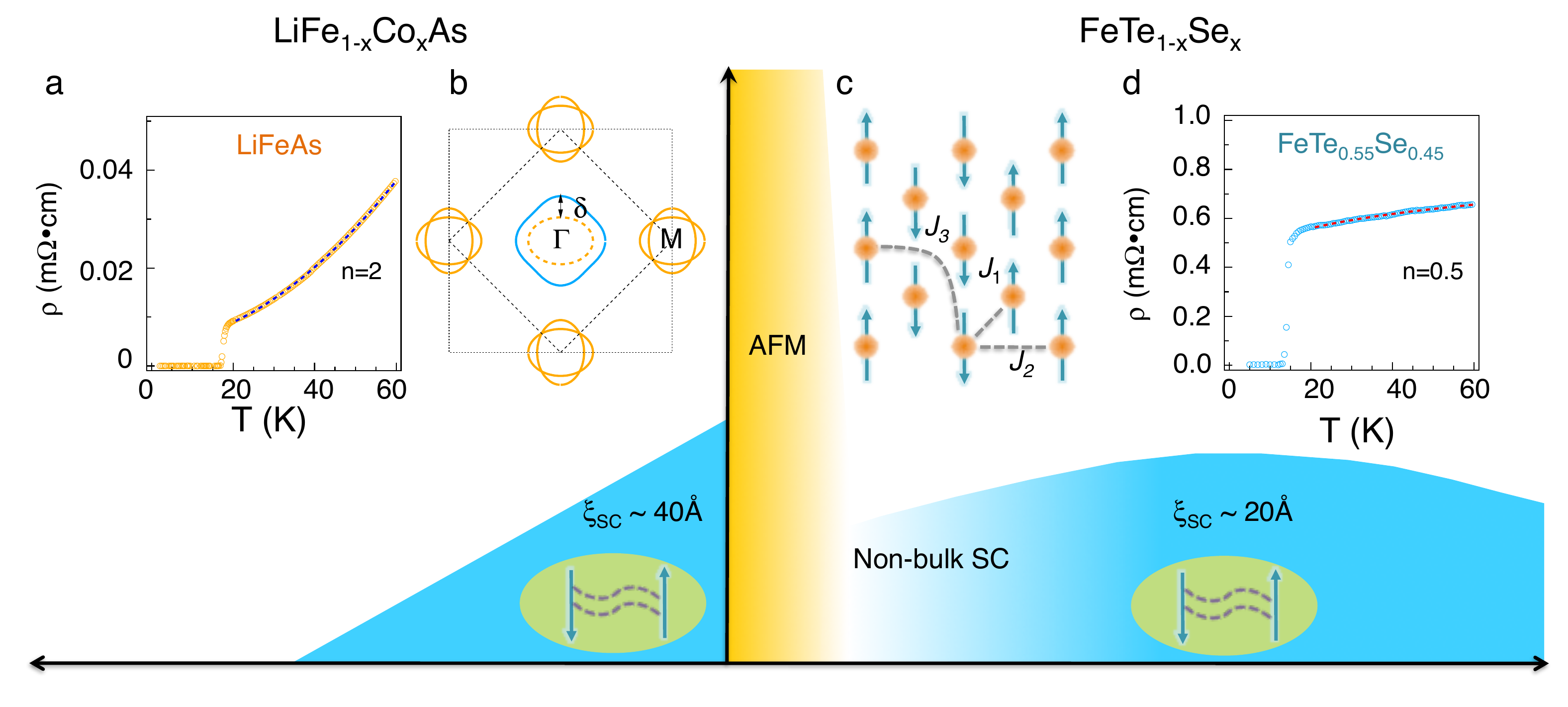}
\caption{(a) Temperature dependent resistivity of \LFA{} shows Fermi liquid behavior up to 60~K. The dashed line is a power function, a+bT$^{n}$, fitting of the data. (b) Experimentally determined FS topology of \LFA{}. The dashed orange ellipse at the $\Gamma$ point is moved from the M point to show the FS size difference, $\delta$, that gives rise the incommensurate low-energy spin excitations \cite{Wang2012}. (c) BC-AFM order of FeTe that is better described by strong coupling $J_{1}$-$J_{2}$-$J_{3}$ model and is not obtained by FS nesting scenario. (d) Temperature dependent resistivity of SC \FTS{} shows a bad metal normal state. The superconducting coherence length, $\xi_{SC}$, of \LFA{} and \FTS{} are 40\AA~and 20\AA, respectively \cite{Allan2012, Yin2015}.}
\label{Fig1}
\end{figure*}

In this study, we choose prototypical FeSCs, \LFA{} and \FTS{}, that have similar SC transition temperature. High-energy resolution ARPES data were recorded at the Institute of Physics, Chinese Academy of Sciences with a Scienta R4000 analyzer. We use the He I$\alpha$ ($h\nu$=21.2~eV) resonance line of an helium discharge lamp. The angular and momentum resolutions were set to 0.2$^{\circ}$ and 2~meV, respectively. All samples were cleaved in situ and measured in a vacuum better than 3$\times10^{-11}$ Torr. Sample orientation and the experimental geometry for the \LFA{} and \FTS{} measurements are the same. Our DFT+DMFT calculations were performed at 116 K within the fully charge self-consistent combination of DFT and embedded dynamical mean field theory (DMFT) \cite{haule_dmft}. The DFT part of these calculations were performed with the WIEN2k package while the DMFT impurity problem was solved by using continous time quantum Monte Carlo (CTQMC) calculations \cite{haule_ctqmc}, with a Hubbard $U = 5.0$ eV and Hund's coupling $J = 0.8$ eV. We use experimental lattice parameters for the calculation of \LFA{} and the averaged anion height to model the \FTS{} alloy \cite{Supp}.

We begin by establishing the distinct normal state electronic coherence of \LFA{} and \FTS{}. As shown in Fig.~\ref{Fig1}a, the pristine \LFA{} has a SC ground state and a Fermi liquid normal state with $T$-quadratic resistivity up to 60 K. The experimentally determined Fermi surfaces (FSs) of \LFA{} are shown in Fig.~\ref{Fig1}b: the mismatch, $\delta$, between the large hole FS at the $\Gamma$ point and the two electron FSs at the M point is found to give rise to incommensurate low-energy spin excitations \cite{Li2016, Qureshi2012, Wang2012, Knolle2012}. In \FTS{}, however, superconductivity is induced by suppressing the bicollinear antiferromagnetic (BC-AFM) phase (Fig.~\ref{Fig1}c). Figure~\ref{Fig1}d shows the resistivity of SC \FTS{}. The normal state resistivity, $\rho_{T_{c}}$=$0.56~m\Omega~cm$, is two orders of magnitude larger than that in \LFA{} and exhibits a saturation behavior in the Mott-Ioffe-Regel limit \cite{Homes2010}, with a mean free path close to the size of the unit-cell. Similar bad metal behavior has also been observed in the pristine FeTe, thus proving that the electronic incoherence is an intrinsic rather than disorder induced property \cite{Dai2014, Ieki2014, Yin2012, Supp}. 

%
\begin{figure*}[tb]
\includegraphics[width=16 cm]{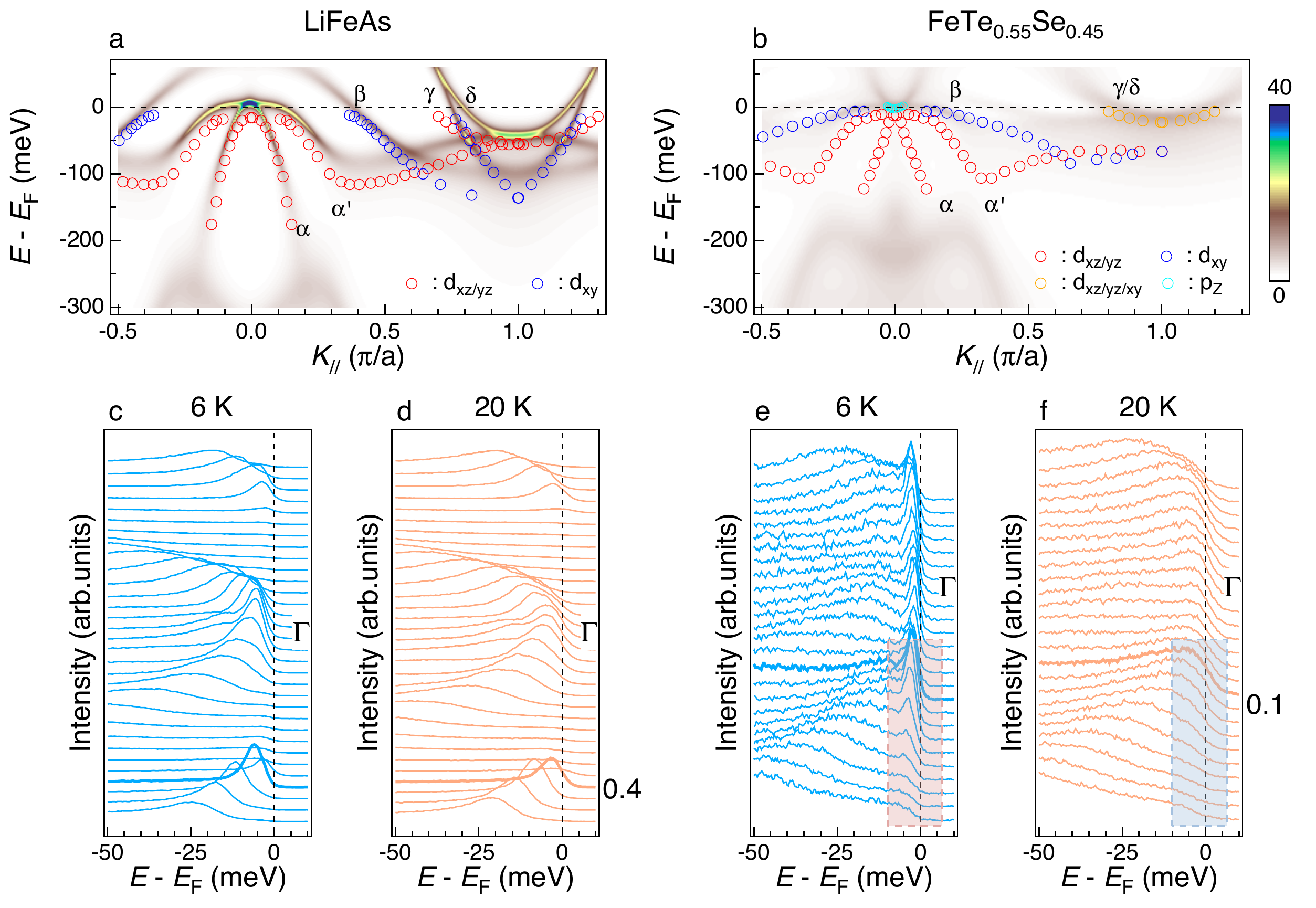}
\caption{(a), (b) DFT+DMFT calculated A(k,$\omega$) without spin-orbit coupling of \LFA{} and \FTS{}, respectively. The colorscales of (a) and (b) are the same. Colored circles are experimentally determined band dispersions along the $\Gamma$-M direction. Orbital contributions of each band are showing in different colors. In the presence of spin-orbital coupling, the $\alpha$' band will be pushed upward and cross $E_{F}$ near the $\Gamma$ point \cite{Johnson2015, Miao2015}. ARPES measured EDCs below and above $T_{c}$ in \LFA{} and \FTS{} are shown in (c), (d) and (e), (f) respectively. The shaded area in (e) and (f) cover the $d_{xy}$ band near $E_{F}$. The thick EDC at 0.4~$\pi/a$ in \LFA{} and 0.1~$\pi/a$ in \FTS{} are corresponding to their $k_{F}^{\beta}$. Due to the intrinsic incoherence of the $\beta$ band in \FTS{}, $k_{F}^{\beta}$ is determined by the minimum gap position in the SC phase and consistent with previous studies \cite{Miao2012, Okazaki2014,Rinotte2017, Ieki2014}.}
\label{Fig2}
\end{figure*}
The different normal state properties between \LFA{} and \FTS{} are indeed captured by our DFT plus dynamic mean field theory (DFT+DMFT) calculations without spin-orbit coupling. Figures~\ref{Fig2}a and b show the DFT+DMFT calculated A(k,$\omega$) superimposed with the ARPES determined band dispersion of \LFA{} and \FTS{}, respectively~\cite{Supp}. As can be seen in these plots, the overall band dispersion agrees quite well with ARPES measurements without any adjustment such as band renormalization and shift. Compared with \LFA{}, the calculated spectral excitation of \FTS{} is broader and more incoherent, thus reflecting its larger scattering rate and smaller $Z_{k}$. These results are in excellent agreement with ARPES measured energy distribution curves (EDCs) in the normal state (T=20~K) as shown in Figs.~\ref{Fig2}d and f. The resolution-limited EDCs near the Fermi level in \LFA{} directly demonstrate the existence of well defined QPs while the linewidth in \FTS{} is significantly broader especially for the most correlated $\beta$ band which, as we show in the light-blue-shaded area of Fig.~\ref{Fig2}f, appears as a weak shoulder on the tail of the $\alpha$' band due to the small $Z_{k}^{\beta}$. In addition, we find that due to the enhanced orbital-selective interaction in \FTS{}, the band width of the $\beta$ band, that is mainly composed of the $d_{xy}$ orbital character, is significantly reduced. This makes \FTS{} close to a semi-metal with the total Fermi energy, $E^{tot}_{F}$, defined as the largest energy difference between the bottom of the electron bands at the M point and the top of the hole bands at the $\Gamma$ point, being 25 meV to be compared with the value of 200 meV in \LFA{}.

\begin{figure*}[tb]
\includegraphics[width=11.6 cm]{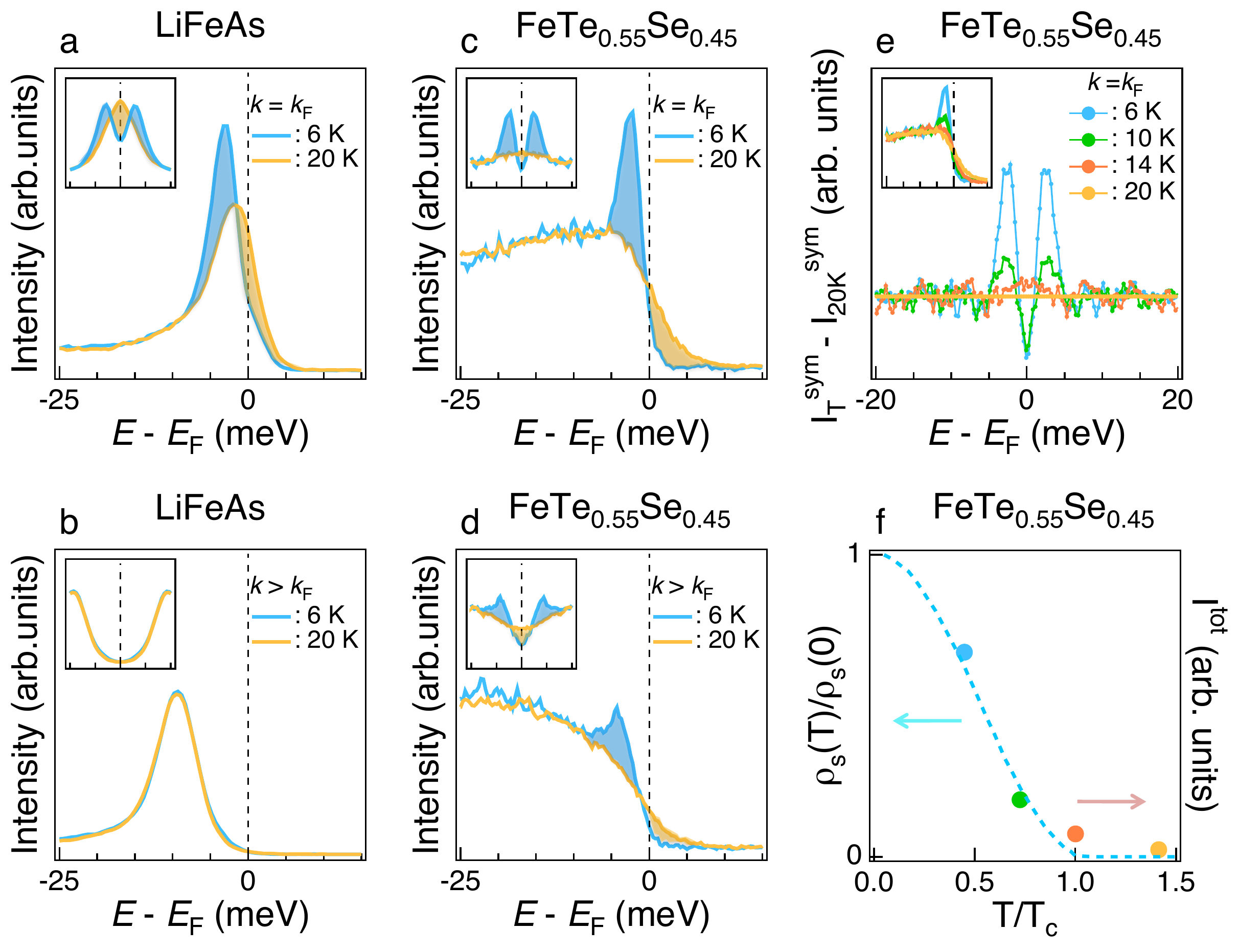}
\caption{(a), (b) EDCs at $k=k_{F}^{\beta}$ and $k>k_{F}^{\beta}$ (0.45$\pi/a$) in \LFA{}. Inset panels show the symmetrized EDCs in (a) and (b). (c), (d) EDCs at $k=k_{F}^{\beta}$ and $k>k_{F}^{\beta}$ (0.2$\pi/a$) in \FTS{} show enhanced total spectral weight in the SC phase. (e) Temperature dependent symmetrized EDCs at $k=k_{F}^{\beta}$. The 20 K data is subtracted from each symmetrized EDCs. Inset shows the temperature dependent raw data at $k=k_{F}^{\beta}$. (f) The integrated intensity of the data in (e) follows the trend of temperature dependent superfluid density in FeTe$_{1-x}$Se$_{x}$ \cite{Kim2010}. The temperature dependent EDCs are normalized by their total counting time.}
\label{Fig3}
\end{figure*}

Having the normal state established, we now explore the corresponding $A(k,\omega)$ response in the SC state. Figures~\ref{Fig2}c and e show the same ARPES EDCs as in Figs.~\ref{Fig2}d and f but now measured in the SC phase (T=6~K). We find that in \LFA{} the resolution limited peaks near $E_{F}$ are shifted to higher binding energies due to the formation of Bogoliubov QPs. In contrast, in \FTS{}, an intense and sharp coherence peak suddenly develops in the SC phase. This contrast is strongest in the shaded areas shown in Figs.~\ref{Fig2}e and f. More strikingly, the SC coherent peaks extend to momenta $k>k_{F}$ on the hole-like $\beta$ band, indicating a non-BCS spectral function \cite{Supp, Feng2000, Ding2001}. To quantitatively compare the ARPES spectra change from the normal to SC state, we show EDCs at $k=k^{\beta}_{F}$ and $k>k^{\beta}_{F}$ of \LFA{} and \FTS{} in Fig.~\ref{Fig3}. In the BCS theory, the SC spectral function is expressed as:
\begin{equation}
A(k,\omega)=\frac{1}{2}\left[\frac{\Gamma_{k}(1+\frac{\xi_{k}}{E_{k}})}{(\omega-E_{k})^{2}+\Gamma_{k}^{2}}+\frac{\Gamma_{k}(1-\frac{\xi_{k}}{E_{k}})}{(\omega+E_{k})^{2}+\Gamma_{k}^{2}}\right]
\label{BCS} 
\end{equation}
\noindent with
\begin{equation}
E_{k}=\sqrt{\xi_{k}^{2}+\Delta_{k}^{2}}
\label{Bog} 
\end{equation}
%

%
%

\noindent where $E_{k}$ and $\xi_{k}$ are the EDC peak position in the SC and normal states, respectively, and $\Delta_{k}$ is the SC gap. In \LFA{}, the change of EDCs is largest near $k_{F}$ and get smaller when $\xi_{k}>>\Delta_{k}$, consistent with Eq.~\ref{BCS} and \ref{Bog}. In addition, we find that the total spectral weight of the symmetrized EDC at $k=k^{\beta}_{F}$ is nearly conserved, which, again, is in agreement with the BCS spectral function. In \FTS{}, however, the change of EDCs is very similar to those observed in the anti-nodal region of cuprates, where the SC coherent peak develops from the incoherent normal state and gains more spectral weight \cite{Fedorov}. In Fig.~\ref{Fig3}e, we symmetrize EDCs at $k_{F}^{\beta}$ at 6~K, 10~K, 14~K and 20~K, and then subtract the 20~K symmetrized intensity. Apparently, the SC coherent spectral weight and the total integrated spectra, $I^{int}$, in $\pm$ 20~meV energy window are continuously increasing as we cool to lower temperature. As shown in Fig.~\ref{Fig3}f, $I^{int}$ indeed tracks the trend of the temperature dependent superfluid density extracted from ref.~\cite{Kim2010}.  

\begin{figure}[tb]
\includegraphics[width=8 cm]{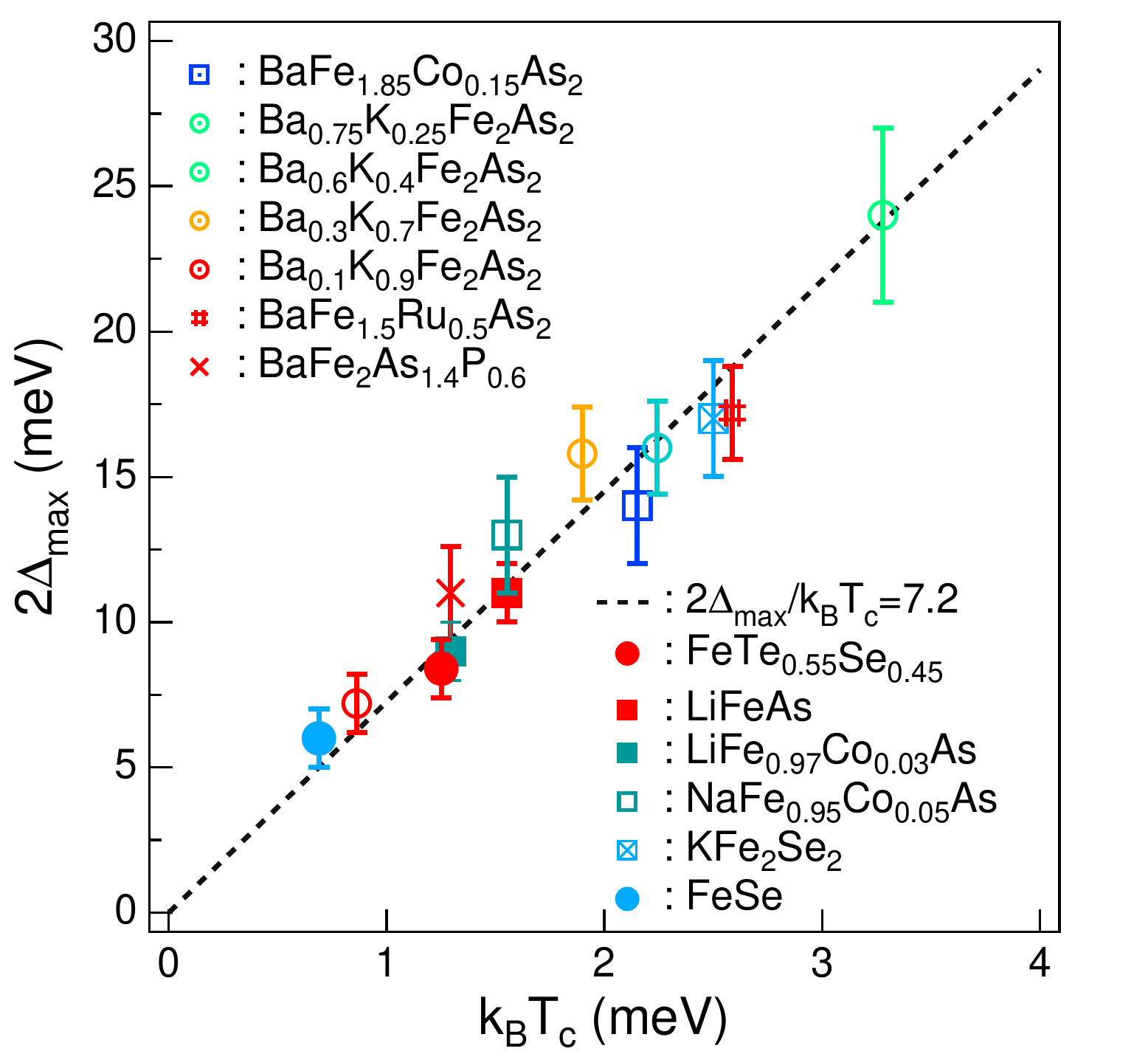}
\caption{Summary of $2\Delta_{SC}^{max}/k_{B}T_{c}$ in various FeSCs that are determined by ARPES \cite{Terashima2009, Nakayama2011, WangXP2011, Xu2013, Liu2011, Miao2012, Miao2016, Ding2008, Xu2013b, ZhangY2012, Liu2018}. The dashed line is a linear function fit of the data points.}
\label{Fig4}
\end{figure}

Despite the dramatic differences on $A(k,\omega)$ and normal state electronic coherence, we find that both \LFA{} and \FTS{} have the same dimensionless quantity $2\Delta_{SC}^{max}(k)/k_{B}T_{c}\sim7.2$, where $\Delta_{SC}^{max}$ is the largest SC gap determined by ARPES. This value is twice larger than that predicted by the BCS theory, confirming the strong pairing nature of these two materials. More intriguingly, as shown in Fig.~\ref{Fig4}, this relation is indeed ubiquitous for all FeSCs covering a wide range of electron filling and distinct FS topologies, dimensionality, impurity level, correlation strength and proximity to quantum criticality. This remarkable universality strongly indicates that all FeSCs share a universal strong coupling pairing mechanism where the largest pairing amplitude, at the lowest order, is decoupled from its electronic coherence. The large impact of the electronic coherence in the normal state on $A(k,\omega)$ in the SC phase is therefore a consequence of the universal and robust SC pairing: the formation of coherent superconductivity, regardless of its microscopic mechanism, reduces the kinetic energy \cite{Hirsch2002} and hence increases the coherent weight of the spectral function. This mechanism is expected to be weak in \LFA{} as the condensed electron pairs mainly originate from the coherent Fermi liquid state \cite{Dai2016}. 

Very recently, the BCS-BEC crossover scenario has been proposed as the possible pairing mechanism for \FTS{} \cite{Okazaki2014, Rinotte2017, Chen2017}, as the SC gap near the $\Gamma$ point is comparable to the $E_{F}$ of the $\beta$ band. As we have already shown in Fig.~2, both \FTS{} and \LFA{} have shallow hole-like FSs near the $\Gamma$ point, and in \LFA{}, the $E_{F}^{\alpha'}$ is even smaller than $\Delta^{\alpha'}$ and can in fact be negative after electron doping \cite{Miao2015}. However, no evidence of BCS-BEC crossover behavior in this system is observed. We point out that in multi-band systems, like the FeSCs, the relevant physical quantity should be $E^{tot}$ that we defined before, rather than the $E_{F}$ for an individual band. Indeed, using the experimentally determined values of $\Delta_{max}$=4.2~meV \cite{Miao2012, Supp} and $E^{tot}$=25~meV, we can nicely reproduce the recently observed Caroli-de Gennes-Martricon states in \FTS{} \cite{Chen2017}. Furthermore, the BCS-BEC crossover scenario is not compatible with the observed universal pairing amplitude with 10 times different $E^{tot}$ in \LFA{} and \FTS{}, and hence cannot be a key ingredient of SC pairing mechanism in FeSCs.

Finally we compare our observations with the cuprate superconductors. While the origin of the electronic interactions and consequently the nature of the normal states are different between the cuprates and FeSCs, their SC response in the charge and spin excitations are remarkably similar. Spin resonance, $\Omega_{res}$, has been observed in both high-$T_{c}$ families \cite{Eschrig2006,DaiPC2015}. The quantity $\Omega_{res}/k_{B}T_{c}\sim5.3$ in the cuprates \cite{Eschrig2006} is larger than $\Omega_{res}/k_{B}T_{c}\sim4.4$ in the FeSCs \cite{WangZY2012}, reflecting a globally larger superconducting energy scale in the cuprate. In addition, the shape of $A(k,\omega)$ in the SC phase is also strongly affected by its normal state $Z_{k}$ in the cuprates, where a BCS-like spectral function is observed near the nodal region and a non-BCS spectral function emerges from the anti-nodal region \cite{Ding2001, Feng2000}. All these similarities suggest unconventional superconductors, including the cuprates and FeSCs, may share a common thread where both the short-ranged AFM spin fluctuations and itinerant carrier are crucial for the pairing mechanism.

%
%

\begin{acknowledgements}
We thank Y. Cao, Y. M. Dai, P. Richard, Y.-L. Wang and W.-G. Yin for useful discussions. H. M., R. D. Z., M.P.M.D and P.D.J. were supported by the Center for Emergent Superconductivity, an Energy Frontier Research Center funded by the U.S. DOE, Office of Basic Energy Sciences. Z.P.Y. was supported by the National Natural Science Foundation of China (Grant No. 11674030), the Fundamental Research Funds for the Central Universities (Grant No.310421113) and the National Key Research and Development Program of China through Contract No. 2016YFA0302300. G.D.G. was supported by the Office of Basic Energy Sciences (BES), U.S. Department of Energy (DOE), through Contract No. de-sc0012704.
\end{acknowledgements}

\bibliography{biblio}

\end{document}